\documentclass[twocolumn,showpacs,preprintnumbers,amsmath,amssymb]{revtex4}
\usepackage{graphicx}
\usepackage{bm}

\newcommand{\ie}{{\it i.e.}}

\def\bra#1{ \langle #1\!\mid }
\def\ket#1{\mid\!#1\rangle}

\def\ave#1{\langle~#1~\rangle}

\begin{document}

\title{The role of inter-well tunneling strength on coherence dynamics of  two-species Bose-Einstein condensates  }

\author{Li-Hua Lu, Xiao-Qiang Xu  and You-Quan Li}
\affiliation{Zhejiang Institute of Modern Physics and Department of Physics,
Zhejiang University, Hangzhou 310027, People's Republic of China}


\begin{abstract}
Coherence dynamics of two-species Bose-Einstein condensates in double wells
is investigated in mean field approximation.
We show that the system can exhibit decoherence phenomena even without
the condensate-environment coupling and
the variation tendency of the degree of coherence depends
on not only the parameters of the system but also the initial states.
We also investigate the time evolution of the degree of coherence for a  Rosen-Zener form of tunneling strength,
and propose a method to get a condensate system with certain degree of coherence through a time-dependent tunneling strength.
\end{abstract}
\received{\today} 


\maketitle
\section{Introduction}

The manipulating Bose-Einstein condensates (BECs)
in double wells provides us a versatile tool
to explore the underlying physics in various nonlinear phenomena
since almost each parameter can be tuned experimentally~\cite{Shin}.
There have been many studies on the fascinating features of nonlinear effect,
such as Rabi oscillation~\cite{Milburn,Jai,LH,Sun}, Josephson oscillation~\cite{Gati,Raghavan,Sun},
self trapping~\cite{Albiez,Likharev,Raghavan,LH,Wang2}, and measure synchronization~\cite{Qiu},
in terms of BECs in double wells.
Because the BECs in double wells can be regarded as a two-level system,
it is also expected either to be employed as a possible qubit
or to simulate certain issues~\cite{Sololovski} in quantum computation and information.
Recently, the effect of decoherence of BECs in double wells
was investigated experimentally by means of interference between BECs~\cite{Schumm,Hffer}
and studied theoretically in terms of single-particle density matrix~\cite{wang}.
In order to exhibit the phenomena of decoherence,
they~\cite{Schumm,Hffer,wang} need to introduce the condensate-environment coupling
because one-species BEC in double wells were merely considered there.
In comparison to one-species BEC system,
the two-species system can exhibit distinct decoherence phenomena
due to the existence of the interspecies interaction.
For example, the degree of coherence in a two-species system
can evolve with time without the application of condensate-environment coupling,
which is useful for one to get a system with desired degree of coherence.
It is therefore worthwhile to study the coherence dynamics of two-species BECs in double wells.

In this paper, we study the effects of the inter-well tunneling strength on the coherence dynamics
for a system of two-species BECs in double wells.
With the help of the reduced single-particle density matrix,
we show that such a system can exhibit decoherence phenomena without condensate-environment coupling.
We also propose an experimental strategy to prepare a BEC system
with any desired degree of coherence through a time-dependent tunneling strength.
In the next section, we model the two-species BEC system and
introduce the reduced single-particle density matrix.
In Sec.~\ref{sec:decoherence}, we study the time evolution of the degree of coherence
for a time-independent inter-well tunneling strength.
In Sec.~\ref{sec:contro}, we investigate the time evolution of the degree of coherence
for a Rosen-Zener form of tunneling strength
and discuss the possibility of preparing a BEC system with any degree of coherence.
Then we briefly give our conclusion in Sec.~\ref{sec:sum}.

\section{Model and reduced density matrix}\label{sec:model}
We consider a two-species Bose-Einstein condensate system confined in double wells.
The Hamiltonian is given by
\begin{eqnarray}\label{eq:hamil}
\hat{H}&=&-J_a(\hat{a}_1^\dagger\hat{a}_2+\textrm{H.c.})-J_b(\hat{b}_1^\dagger\hat{b}_2
+\textrm{H.c.})+
\sum_{i}{\tilde{U}_{ab}\hat{n}_{a i}\hat{n}_{b i}}\nonumber\\
&&+\sum_{ i}{\frac{1}{2}\tilde{U}_{aa}\hat{n}_{a i}\hat{n}_{a i}}
+\sum_{ i}{\frac{1}{2}\tilde{U}_{bb}\hat{n}_{b i}\hat{n}_{b i}},
\end{eqnarray}
where $\hat{a}_i^\dagger$ ($\hat{b}_i^\dagger$) and $\hat{a}_i$ ($\hat{b}_i$) creates and annihilates a bosonic atom of species
$a$ ($b$) in the $i$th well, respectively;  $\hat{n}_{a i}=\hat{a}_i^\dagger\hat{a}_i$  ($\hat{n}_{b i}=\hat{b}_i^\dagger\hat{b}_i$) denotes the particle number operator of species $a$ ($b$). Here the parameters $J_a$ and $J_b$ denote the tunneling strengths of species $a$ and $b$ between the two wells, $\tilde{U}_{aa}$ and $\tilde{U}_{bb}$ are the intraspecies interaction strengths, and $\tilde{U}_{ab}$ is the interspecies interaction strength. The Hamiltonian (\ref{eq:hamil}) can describe a BEC mixture
confined in a double well potential consisting of different atoms, or different isotopes, or different hyperfine states of the same kind of atom. The coherence dynamics of the above model  has not been investigated although its dynamical properties, like  Josephson oscillation, stability and measure synchronization etc., have been  studied in earlier works \cite{Qiu,Sun,xiq}. Whereas, we know that the role of coherence of the system is very important since  the first obstacle attempted to be avoided is the decoherence when a condensate
in double wells is expected to be employed as a qubit. So  the coherence dynamics of
the model~(\ref{eq:hamil}) is worthy of study.

 As we know, under the semiclassical limit~\cite{Likharev,Milburn,Leggett,Gati}, the dynamics of this system
 is conventionally  studied in mean-field approach by replacing the expectation values of annihilators with
 complex numbers, \ie, $\ave{\hat{a}_ i}=\tilde{a}_ i$ and $\ave{\hat{b}_ i}=\tilde{b}_i$. With the help of Heisenberg
 equation of motion for operators,
 one can easily get the dynamical equations for $\tilde{a}_i$ and $\tilde{b}_i$. These equations guarantee
 the conservation law $|\tilde{a}_1|^2+|\tilde{a}_2|^2=N/2$ and $|\tilde{b}_1|^2+|\tilde{b}_2|^2=N/2$ with $N$
 being the total particle number of species $a$ and $b$.
 To simplify the
calculation, one usually assumes
 $a_i=\tilde{a}_i/\sqrt{N}$ and $b_i=\tilde{b}_i/\sqrt{N}$. Then  the aforementioned
dynamical equations for $\tilde{a}_i$ and $\tilde{b}_i$ can be rewritten as,
 \begin{equation}\label{eq:dynapure}
 i\frac{d}{dt}\ket{\psi}=H_{\textrm{eff}}
  \ket{\psi}
   \end{equation}
 with
 \begin{widetext}
 \begin{equation}\label{eq:haeff}
 H_{\textrm{eff}}=\left(
   \begin{array}{cccc}
     U_a|a_1|^2+U_{ab}|b_1|^2 & -J_a & 0 & 0 \\
     -J_a & U_a|a_2|^2+U_{ab}|b_2|^2 & 0 & 0 \\
     0 & 0 & U_b|b_1|^2+U_{ab}|a_1|^2 & -J_b \\
     0 & 0 & -J_b & U_b|b_2|^2+U_{ab}|a_2|^2 \\
   \end{array}
 \right),
 \end{equation}
 \end{widetext}
where $U_a=\tilde{U}_{aa}N$, $U_b=\tilde{U}_{bb}N$, and
$U_{ab}=\tilde{U}_{ab}N$.
Here the wave function $\ket{\psi}$ refers to
\begin{equation}
\ket{\psi}=\left(\begin{array}{cccc}
  a_1, & a_2, & b_1, & b_2 \\
    \end{array}
         \right)^\textrm{T}
   =\sum_{\sigma,i}\ket{\sigma}\otimes\ket{i},
\end{equation}
where the state $\ket{\sigma}=\ket{a}$ or $\ket{b}$ specifies the two different species while $\ket{i}=
\ket{1}$ or $\ket{2}$ specifies the two wells. Note that the dynamical properties of the system can be determined by   Eq.~(\ref{eq:dynapure})
if the system is in a completely
coherent state (\ie, a pure state). Whereas, if the system is in a mixed state,
the equation~(\ref{eq:dynapure})
becomes insufficient.

In order to  study the coherence dynamics of the system, we introduce the single-particle
density matrix $\hat{\rho}=\ket{\psi}\bra{\psi}$
 whose elements are
$\rho_{\mu\nu}(\mu, \nu=1,2,3,4)$.  From this definition, we can see that, as a
$4 \times 4$ matrix, $\hat{\rho}$ describes a pure state. Their diagonal elements $\rho_{11}$ ($\rho_{22}$) and $\rho_{33}$ ($\rho_{44}$) represent the population of species $a$ ($b$) in the first and second well, respectively.
In this paper, we only focus on the  distribution of the total particle numbers  in the two  wells but not distinguish the particle
species, so the system can be described by the reduced density matrix
\begin{equation}\label{eq:reducematrix}
\hat{\rho}^r=\sum_\sigma\bra{\sigma}\hat{\rho}\ket{\sigma},
\end{equation}
whose elements are $\rho_{11}^r=\rho_{11}+\rho_{33}$, $\rho_{12}^r=\rho_{12}+\rho_{34}$, $\rho_{21}^r=\rho_{21}+\rho_{43}$, and $\rho_{22}^r=\rho_{22}+\rho_{44}$.
Clearly, the matrix $\hat{\rho}^r$ can describe a mixed state. Its diagonal elements $\rho_{11}^r$ and $\rho_{22}^r$ represent the total population probability in
the first and second well, respectively. To
investigate the coherence dynamics of the system, we can
introduce the definition of degree of coherence
according to Ref.~\cite{Leggett,LH},
\begin{equation}\label{eq:coherence}
\eta=\textrm{Tr}(\hat{\rho}^r)^2-1.
\end{equation}
From Eq.~(\ref{eq:reducematrix}) and Eq.~(\ref{eq:coherence}), we can see that
the time evolution of the degree of coherence depends on that of the elements $\rho_{\mu\nu}$ of the
single-particle density matrix $\hat{\rho}$. Since $\hat{\rho}$ describes a pure state, the  time evolution of $\rho_{\mu\nu}$ can be determined by Eq.~(\ref{eq:dynapure}). Thus one can solve Eq.~(\ref{eq:dynapure}) to get the evolution  of $a_i$ and $b_i$ firstly,  and then  gives  the time evolution of $\rho_{\mu\nu}$ according to  the definition of $\hat{\rho}=\ket{\psi}\bra{\psi}$.  Then in the following sections, we will study the coherence dynamics of the system with the help of Eq.~(\ref{eq:dynapure}).
Note that in the following calculations, we take $J_a=J_b=J$ for
simplicity without losing the generality.

\section{Evolution of  degree of coherence }\label{sec:decoherence}
\begin{figure}[ht]
\includegraphics[width=90mm]{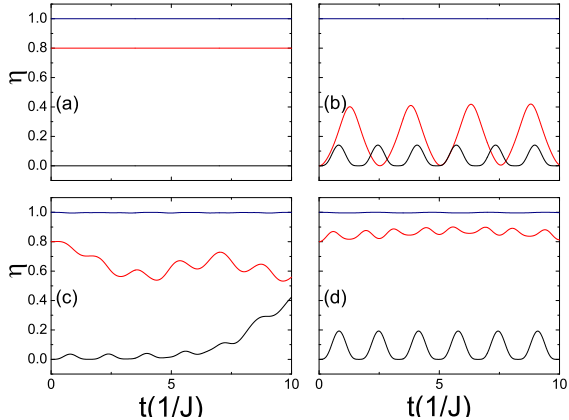}
\caption{\label{fig:coherence}(Color online) Time evolution of the degree of coherence for different
initial states.
The  parameters are
$U_a=U_b=U_{ab}=3J$ (a), $U_a=U_b=3J$ and $U_{ab}=6J$ (b),
$U_{a}=3J, U_{b}=2J$ and $U_{ab}=J$ (c), and $U_{a}=3J, U_{b}=2J$ and $U_{ab}=6J$ (d).
 }
\end{figure}

We know that the degree of coherence does not change for an isolated BEC system~\cite{LH}. To
study the effect of decoherence of BEC system, several authors considered the condensate-environment
coupling~\cite{wang,Kho}. Whereas, we will show that the degree of coherence of the two-species
BEC system in double wells can still change with time even without  the condensate-environment coupling.
Here we consider the case of time-independent inter-well tunneling strength, \ie, $J$ is a constant in the calculation. Due to the fact that Eq.~(\ref{eq:dynapure}) can not be analytically solved, we solve Eq.~(\ref{eq:dynapure}) numerically to get the time evolution of $a_i$ and $b_i$.  Then we give the time evolution of the degree of coherence with the help of the definition of $\hat{\rho}$ and $\hat{\rho}_r$ and Eq.~(\ref{eq:coherence}). The corresponding results are summarized in
Fig.~\ref{fig:coherence}.

In Fig.~\ref{fig:coherence}, we plot the time evolution of the degree of coherence for
different  initial states and parameters. The initial states are $a_1=\sqrt{1/2}, a_2=0, b_1=\sqrt{1/2}$
and $b_2=0$ for the blue (top) line, $a_1=\sqrt{2/5}, a_2=\sqrt{1/10}, b_1=\sqrt{1/2}$ and $b_2=0$ for
the red (middle) line, and $a_1=\sqrt{1/2}, a_2=0, b_1=0$ and $b_2=\sqrt{1/2}$
for the black (bottom) line.
From the Fig.~\ref{fig:coherence} (a), we can find that the degree of coherence  of the two-species BEC
system in double wells does not change for any initial states when $U_{a}=U_b=U_{ab}$.
That  can be easily  understood because  this two-species system is equivalent to
the one-species system once $U_a=U_b=U_{ab}$. And according to Ref.~(\cite{LH}), the degree
of coherence of the one-species BEC system in double wells does not change
without the condensate-environment coupling. So the result shown in Fig.~\ref{fig:coherence} (a)
is reasonable. The figure~\ref{fig:coherence} (b) shows that when  $U_a=U_b\neq U_{ab}$, the degree of coherence does not change for the  case  of $a_i=b_i$ at the initial time, \ie, the initial population distributions
of species $a$ and $b$ are the same,
but that changes for the other initial states. Note that for the
case of $U_a=U_b\neq U_{ab}$,  the time evolutions of species $a$
and $b$ are the same if $a_i=b_i$ at the initial time. So the degree of coherence does not change due to the two species  having the same symmetry.
Additionally, from the Fig.~\ref{fig:coherence} (c) and (d), we can find that the degree of coherence
changes for any initial states for the case of $U_a\neq U_b\neq U_{ab}$. Comparing
Fig.~\ref{fig:coherence} (c) and (d), we can see that for the same initial state, the
variation tendency of the degree of coherence depends on the parameters of the system.
In summary, whether the degree of
coherence  can change with time depends on both
the parameters and the initial states. So does the variation tendency of the degree of coherence. This fact indicates that one can
control the degree of coherence without introducing the effect of environment, which will
be discussed in the next section.

\section{controlling the degree of coherence }\label{sec:contro}

 In previous section, we show that the time evolution of the degree of coherence depends on the initial states and the parameters of the system, so one can change it by varying  either the initial states or the parameters of the system. In the following, we will show how to  control the degree of coherence through  a Rosen-Zener form of $J$,
\begin{equation}
J=J_0\sin^2{\omega t}  \quad\quad(0\leq t\leq \pi/\omega),
\end{equation}
\ie, $J$ increases from zero to its maximum value $J_0$ and then decreases to zero again in the end of the calculation.  In the numerical calculation in this section, we take the initial state $a_1=\sqrt{0.4}, a_2=\sqrt{0.1}, b_1=\sqrt{0.5}$, and $b_2=0$.
\begin{figure}[ht]
\includegraphics[width=50mm]{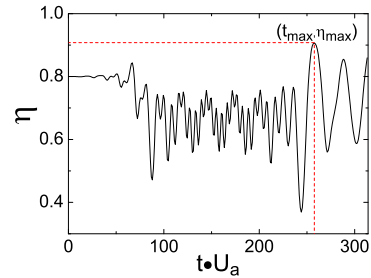}
\caption{\label{fig:evoRosen}(Color online) Time evolution of the degree of coherence for the Rosen-Zener form of  inter-well tunneling strength.
The  parameters are
 $\displaystyle U_b=\frac{2}{3}U_a,  U_{ab}=\frac{1}{3}U_a, J_0=\frac{1}{3}U_a, \omega=0.01U_a$.
 }
\end{figure}
\begin{figure}[ht]
\includegraphics[width=90mm]{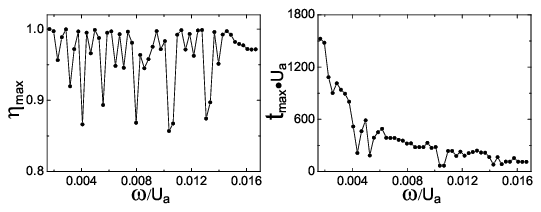}
\caption{\label{fig:omega} The dependence of the maximum value of the degree of coherence (left panel) and the corresponding time (right panel)  on the period of the inter-well tunneling strength.
The  parameters are
$\displaystyle U_b=\frac{2}{3}U_a,  U_{ab}=\frac{1}{3}U_a, J_0=\frac{1}{3}U_a$.
 }
\end{figure}

\begin{figure}[ht]
\includegraphics[width=90mm]{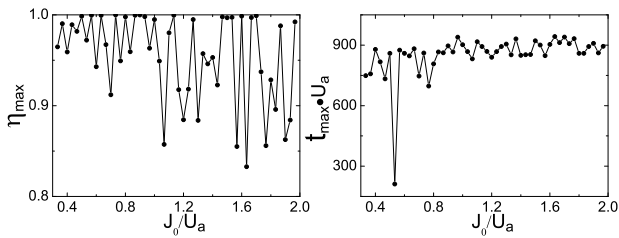}
\caption{\label{fig:tuj} The dependence of the maximum value of the degree of coherence (left panel) and the corresponding time (right panel)  on the maximum value of the inter-well tunneling strength.
The  parameters are
$\displaystyle U_b=\frac{2}{3}U_a,  U_{ab}=\frac{1}{3}U_a, J_0=\frac{1}{3}U_a$.
 }
\end{figure}

 The time evolution of  the degree of coherence for the Rosen-Zener form of inter-well tunneling strength is plotted in Fig.~\ref{fig:evoRosen}. We can see that the degree of coherence  changes with time and  can reach the maximum value $\eta_\textrm{max}$ at time $t=t_\textrm{max}$. Note that the values of  $\eta_\textrm{max}$ and $t=t_\textrm{max}$ depend on both the period  $\omega$ and the maximum value $J_0$ of the inter-well tunneling strength.
In Fig.~\ref{fig:omega}, we plot  the dependence of  $\eta_\textrm{max}$ and $t_\textrm{max}$ on $\omega$ for a fixed $J_0$ in the left and right panel, respectively. Meanwhile, we plot the dependence of $\eta_\textrm{max}$ and  $t_\textrm{max}$ on  $J_0$ for a fixed $\omega$ in Fig.~\ref{fig:tuj}.

From Fig.~\ref{fig:omega} and Fig.~\ref{fig:tuj}, we can see that  the values of $\omega$ and $J_0$ affect the maximum value of the degree of coherence sensitively. In order  to get a system with large value of  degree of coherence, one must choose the form of $J$ with suitable period $\omega$ and maximum value $J_0$. Note that although one can get a system with large value of degree of coherence through tuning the value of $\omega$ and $J_0$, it is difficult to control the time evolution of the degree of coherence  due to the fact that the degree of coherence does not evolve periodically, which can be confirmed by Fig.~\ref{fig:evoRosen}. In order to overcome the aforementioned  problem, we consider  the following form of $J$
\begin{equation}\label{eq:Jform}
J=\begin{cases}
    J_0\sin^2{\omega t} \quad\quad(0\leq t\leq t_\textrm{max})\\
       0 \quad\quad\quad\quad\quad\quad\quad(  t > t_\textrm{max}),
\end{cases}
\end{equation}
where $t_\textrm{max}$ can be obtained from Fig.~\ref{fig:omega} and Fig.~\ref{fig:tuj}.

\begin{figure}[ht]
\includegraphics[width=50mm]{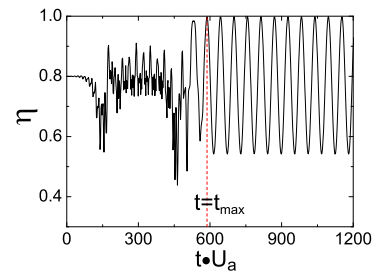}
\caption{\label{fig:evo}(Color online) Time evolution of the degree of coherence for the  inter-well tunneling strength given in Eq.~(\ref{eq:Jform}).
The  parameters are
$\displaystyle U_b=\frac{2}{3}U_a,  U_{ab}=\frac{1}{3}U_a, J_0=\frac{1}{3}U_a, \omega=0.005U_a$.
 }
\end{figure}

For the form of $J$ given in Eq.~(\ref{eq:Jform}), the time evolution of the degree of coherence is plotted in Fig.~\ref{fig:evo}. From this figure, we can find that the degree of coherence reaches its maximum value which is determined by the values of $\omega$ and $J_0$,  and then
oscillates periodically at the following time. Since Eq.~(\ref{eq:dynapure}) can be analytically solved for the case of $J=0$, we can give the oscillation period  in analytical method.
Solving Eq.~(\ref{eq:dynapure}), we obtain $a_i(t)=a_i(t_\textrm{max})\exp{(-iA_it)}$, and $b_i(t)=b_i(t_\textrm{max})\exp{(-iB_it)}$ where $A_i=U_a|a_i(t_\textrm{max})|^2+U_{ab}|b_i(t_\textrm{max})|^2$,
and  $B_i=U_b|b_i(t_\textrm{max})|^2+U_{ab}|a_i(t_\textrm{max})|^2$. Substituting the expressions of $a_i$ and $b_i$ into the Eq.~(\ref{eq:coherence}),  we can get the oscillation period of the degree of coherence $ T=2\pi/|A_2-A_1+B_1-B_2|$. For the parameters taken in Fig.~\ref{fig:evo}, we find $TU_a=56.5$ which is consistent with the numerical result. From the above discussion, we know that the degree of coherence oscillates periodically after time $t_\textrm{max}$, and both the maximum value of the degree of coherence and the oscillation period can be changed by tuning the values of $\omega$  and $J_0$. Once the values of  $\omega$ and $J_0$ are fixed, the time evolution of the degree of coherence is well defined, so that we can know the degree of coherence of the system at any time. Then one can easily get a system with any desired degree of coherence through controlling the evolution time.

\section{Summary}\label{sec:sum}

In the above, we investigated the coherence dynamics for a system of two-species Bose-Einstein condensate in double wells.
In mean field approximation, we studied the influence of the inter-well tunneling strength on the coherence features with the help of the reduced single-particle density matrix.
Since we need not distinguish particle species in the system,
we only focused on the distribution of the total particle numbers in the two wells,
which can be described by a 2$\times$2 reduced density matrix.
After studying the time evolution of the degree of coherence for
a time-independent inter-well tunneling strength,
we found that the degree of coherence of the two-species BEC system
changes for some parameters and initial states even
without the condensate-environment coupling, which differs from the case of Refs.~\cite{wang,Kho}.
Motivated by the fact that the variation tendency of the degree of coherence
depends on both the parameters and the initial states of the system,
we considered a system with a Rosen-Zener form of inter-well tunneling strength
to control the degree of coherence.
Although its tendency is not periodical, the degree of coherence can reach a maximum value $\eta_{\textrm{max}}$ at time $t_{\textrm{max}}$
which are dependent of the period $\omega$ and the maximum value $J_0$ of the inter-well tunneling strength. The dependence of $\eta_\textrm{max}$ and $t_\textrm{max}$ on $\omega$ and $J_0$
we obtained is helpful for one to get a system with large value of degree of coherence
utilizing a Rosen-Zener form of inter-well tunneling strength.
We also gave a useful form of the inter-well tunneling strength
for one to easily get a system with any degree of coherence
by controlling the evolution time.

The work is supported by NSFC under Grant No. 10874149 and No. 11074216.

\end{document}